%% file: Template.tex
\documentclass{article}
\usepackage{spconf,amsmath,graphicx}
\usepackage{amsfonts}
\usepackage{mathtools}

\usepackage{xcolor}
\usepackage{cite}
\usepackage{tikz}
\usepackage{siunitx}
\usepackage{subfigure}
\usepackage{balance}

\usepackage{pgfplots}%
    \usepgfplotslibrary{units} 
    \pgfplotsset{compat=1.9}
\makeatletter
\pgfplotsset{
unit code/.code 2 args=
    \begingroup
    \protected@edef\x{\endgroup\si{#2}}\x
}

\usepackage{bibspacing}
\def\x{\boldsymbol{x}_{q}^{\prime}}
\def\xu{\hat{\boldsymbol{x}}_{q}^{\prime}}
\def\xlu{\hat{\boldsymbol{x}}_{\ell}}
\def\xl{\boldsymbol{x}_{\ell}}

\def\balpha{\boldsymbol{\alpha}_N\left(\hat{\boldsymbol{x}}_{\ell}\right)}
\def\talpha{\tilde{\alpha}_{nm}\left(\hat{\boldsymbol{x}}_{\ell}\right)}

\def\balphad{\boldsymbol{\alpha}_N\left(\hat{\boldsymbol{x}}_{\check{\ell}}\right)}
\def\talphad{\tilde{\alpha}_{nm}\left(\hat{\boldsymbol{x}}_{\check{\ell}}\right)}

\def\An{\overline{\mathbf{A}}_\tau}
\def\hbalpha{\hat{\boldsymbol{\alpha}}_N}
\def\balphan{\|\boldsymbol{\alpha}_N\|}
\def\PD{\operatorname{PD}}
\def\RMSE{\operatorname{RMSE}}
\def\figSize{5cm}

\usepackage{accents}
\newlength{\dhatheight}
\newcommand{\doublehat}[1]{%
    \settoheight{\dhatheight}{\ensuremath{\hat{#1}}}%
    \addtolength{\dhatheight}{-0.35ex}%
    \hat{\vphantom{\rule{1pt}{\dhatheight}}%
    \smash{\hat{#1}}}}

\title{ACOUSTIC SOURCE LOCALIZATION IN THE SPHERICAL HARMONICS DOMAIN EXPLOITING LOW-RANK APPROXIMATIONS}
\name{Maximo Cobos\sthanks{\scriptsize{Mobility Grant PRX21/00174 from Ministerio de Universidades of Spain funded partially this work. Thanks for the additional support provided by Grant RTI2018-097045-B-C21 funded by MCIN/AEI/10.13039/501100011033 and by “ERDF A way of making Europe". Grant TED2021-131003B-C21 funded by MCIN/AEI/10.13039/501100011033 and by the “EU Union NextGenerationEU/PRTR”. The authors acknowledge also the Artemisa computer resources funded by the EU ERDF and Comunitat Valenciana, and the technical support of IFIC (CSIC-UV).}}, Mirco Pezzoli$^\dagger$, Fabio Antonacci$^\dagger$, Augusto Sarti$^\dagger$}
\address{Departament d'Inform\'atica, Universitat de Val\`encia, Valencia, Spain\\
$^\dagger$Dipartimento di Elettronica, Informazione e Bioingegneria, Politecnico di Milano, Milan, Italy}

\begin{document}
\ninept
\maketitle%
\begin{abstract}%
Acoustic signal processing in the spherical harmonics domain (SHD) is an active research area that exploits the signals acquired by higher order microphone arrays. A very important task is that concerning the localization of active sound sources. In this paper, we propose a simple yet effective method to localize prominent acoustic sources in adverse acoustic scenarios. By using a proper normalization and arrangement of the estimated spherical harmonic coefficients, we exploit low-rank approximations to estimate the far field modal directional pattern of the dominant source at each time-frame. The experiments confirm the validity of the proposed approach, with superior performance compared to other recent SHD-based approaches.
\end{abstract}%
\section{Introduction}\label{sec:intro}%
The localization of acoustic sources is a well-known problem in the field of acoustic signal processing \cite{cobos2017survey} that typically concerns the identification of the so-called direction of arrival (DOA) from a multichannel acquisition. 
The DOA information is essential in a great variety of applications such as source separation \cite{gannot2017consolidated, markovic2016extraction} or sound field reconstruction \cite{pezzoli2020parametric,cobos2022overview}. 

A popular class of source localization approaches is based on a  beamformer-like operation, such as the steered response power (SRP) \cite{yao2002maximum} and its variant SRP-phase transform (SRP-PHAT) \cite{dibiase2001robust, cobos2010modified}.
These methods localize the source exploring the whole space of directions while looking for areas where the response power is maximized.
A second important class of approaches is represented by subspace methods \cite{benesty2000adaptive, stoica1989music, jo2018direction}. 
In general, subspace models exploit the decomposition of the spatial covariance matrix (SCM) of the multichannel data in order to identify the source components.
The multiple signal classification (MUSIC) \cite{stoica1989music} algorithm represents a  popular technique of this category, due to its inherent simplicity and effective performance. In practice, MUSIC computes a pseudospectrum over the possible DOAs through the nullspace of the noise eigenvectors of the SCM. The maxima in the pseudospectrum correspond to the estimated DOAs of the sources.

The increasing availability of a high number of sensors in microphone arrays raised the adoption of sound field representations \cite{markovic2013soundfield,williams1999fourier} in order to exploit their characteristics in different applications \cite{pezzoli2021ray,mitsufuji2020multichannel,pezzoli2022sparsity,fahim2017sound} including source localization \cite{bianchi_ray_2016,abhayapala2003coherent,teutsch2005eigen}.
In this context, the sound field decomposition in terms of spherical harmonics (SH) has been widely adopted since it enables the decoupling of frequency-dependent and direction-dependent components of the acoustic field.
It follows that many source localization techniques have been adapted to the SH-domain (SHD), e.g., SHD-MUSIC \cite{abhayapala2003coherent} and EB-ESPRIT \cite{teutsch2005eigen}, showing improved localization performance.

Recently, in \cite{hu2019sound} the relative harmonic coefficients (RHC) were proposed as a useful feature for source localization in the SHD.
The RHCs can be thought of as the SHD counterpart of the relative transfer functions (RTF), containing the DOA information while being independent from the source signal and robust to noise.
Different approaches exploiting RHC were published in the literature, including solutions based on grid search \cite{hu2020unsupervised}, gradient descent \cite{hu2021evaluation} and Gaussian process \cite{hu2019sound}. 
In \cite{hu2021decoupled, hu2022decoupled}, computationally efficient approaches employing a search decouple on azimuth and elevation were introduced, while \cite{hu2022closed} provides a closed-form solution limited to \textit{first}-order SHD.
Inspired by the effectiveness of RTF and RHC as features for localization, in \cite{hu2020multiple} the authors introduced the relative sound pressure MUSIC (RMUSIC) and its SHD version (SHD-RMUSIC) showing improved performance with respect to the traditional methods.

In this work, we propose a novel DOA estimation approach that exploits low-rank signal approximations in the SHD, referred to as SHD-LRA.
Similarly to other SHD-based solutions, we exploit this representation in order to estimate the direction-dependent components that identify the source location. 
Differently from RHC models, the proposed technique works directly on the coefficients of the spherical-harmonics-transformed array signals.
Therefore, it does not rely on any time-averaging operation typically required by other methods.
In particular, similarly to subspace-based solutions, we rely on a low rank approximation of the data. 
However, the decomposition is not applied on the SCM as in MUSIC-like methods, but on the SHD signal, exploiting the properties of the so-called modal directional pattern (MDP). 
MDPs are defined as frequency-independent far-field model of the sources hence, they are analytically known.
The proposed technique consists of three main steps: (1) the acquired SHD coefficients are normalized in order to match the ideal MDP norm under a single source assumption (W-disjoint orthogonality assumption); (2) the low rank approximation of the normalized SHD data is performed through the singular value decomposition (SVD) of the normalized coefficients identifying the MDP of the primary source; (3) the DOA estimate is retrieved by pattern matching over a set of pre-computed MDP prototypes.

We compare the performance of the proposed model with respect to recent low-rank solutions that exploit the RTF model, namely RMUSIC and SHD-MUSIC.
Results show that the proposed localization method provide a robust performance with high reverberation and low SNR, outperforming both RMUSIC and SHD-MUSIC.
The rest of the paper is organized as follows.
In Sec.~\ref{sec:problem_formulation}, the problem of source localization in the SHD is introduced and we provide the definition of the MDPs.
Sec.~\ref{sec:method} describes in details the proposed method SHD-LRA based on low-rank approximation of SHD signals.
In Sec.~\ref{sec:experiments}, we provide the validation of SHD-LRA and its performance is compared with respect to the reference techniques.
Finally, in Sec.~\ref{sec:conclusion} we draw conclusions and propose future developments.
\section{PROBLEM FORMULATION}%
\label{sec:problem_formulation}%
The measured sound pressure $P$ corresponding to a continuous sound field on a sphere of radius $R$ can be decomposed
using the SH basis functions as \cite{williams1999fourier}
\begin{equation}
P\left(\x, k\right)=\sum_{n m}^{\infty} \alpha_{n m}(k) b_{n}\left(k R\right) Y_{n m}\left(\xu\right),
\label{eq:pressure}
\end{equation}
where $\sum_{n m}^{(\cdot)} \equiv \sum_{n=0}^{(\cdot)} \sum_{m=-n}^{n} \alpha_{n m}(k)\in\mathbb{C}$ is the sound field coefficient of order $n$ and degree $m$. The position vector is $\x \equiv\left(R, \xu\right)$, with the unit vector $\xu \equiv\left(\theta_{q}^{\prime}, \phi_{q}^{\prime}\right)$ indicating the elevation and azimuth of $\x$. The infinite summation of Eq.~\eqref{eq:pressure} is often truncated at the sound field order $N=\left\lceil k R\right\rceil$ \cite{ward2001reproduction}, where $k=\frac{2\pi f}{c}$ for a frequency $f$ and propagation speed $c$, and $\lceil\cdot\rceil$ denotes the ceiling operation due to the high-pass nature of the higher-order Bessel functions. Then, for a maximum order $N$ there are $C = (N+1)^2$ coefficients.
The complex SH basis function $Y_{n m}(\cdot)$ is defined as
{\small \begin{equation}
Y_{n m}\left(\xu\right)=\sqrt{\frac{(2 n+1)}{4 \pi} \frac{(n-|m|) !}{(n+|m|) !}} \mathcal{P}_{n|m|}\left(\cos \theta_{q}^{\prime}\right) e^{i m \phi_{q}^{\prime}}
\end{equation}}
where $|\cdot|$ denotes absolute value, $(\cdot)$! represents factorial, $\mathcal{P}_{n|m|}(\cdot)$ is an associated Legendre polynomial, and $i~=~\sqrt{-1}$. Furthermore, the dependency on array radius comes through the function $b_{n}(\cdot)$ which is defined as
\begin{equation}
b_{n}(\xi)= \begin{cases}j_{n}(\xi) & \text { for an open array } \\
j_{n}(\xi)-\frac{j_{n}^{\prime}(\xi)}{h_{n}^{\prime}(\xi)} h_{n}(\xi) & \text { for a rigid spherical array }\end{cases}
\end{equation}
where $h_n(\cdot)$ and $j_n(\cdot)$ are the $n$th order spherical Hankel and Bessel functions of the first kind, respectively.

\begin{figure}%
    \centering%
    \includegraphics[width=0.55\columnwidth]{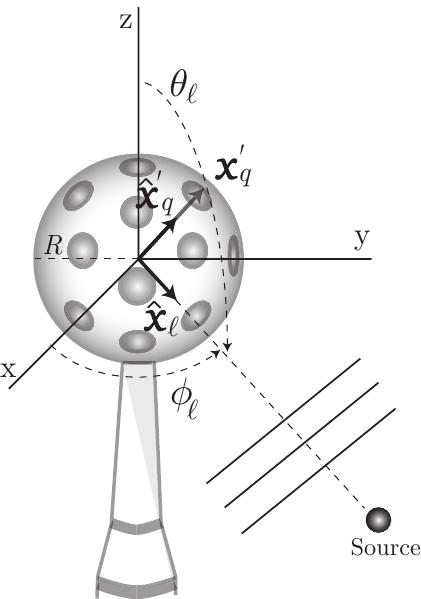}%
    \caption{Reference geometry for the proposed method.}\vspace{-1.2em}
    \label{fig:setup}
\end{figure}

Given the far-field approximation of the Green's function, the SH coefficients considering only the direct propagation paths from $L$ sources are given by 
\begin{equation}
  \alpha_{n m}(k)=4 \pi i^{n} \sum_{\ell=1}^{L} S_{\ell}(k)G_{\ell}^{(d)}(k) Y_{n m}^{*}\left(\xlu\right),  
\end{equation}
where $S_{\ell}(k)$ is the $\ell$th source signal and $G_{\ell}^{(d)}(k)$ represents the direct path gain between the origin and the $\ell$th source location $\xl$, with DOA vector $\xlu$. Note that the coefficients can be compactly written as:
\begin{equation}
\alpha_{n m}(k) = \sum_{\ell=1}^{L}\talpha \tilde{S}_{\ell}(k)
\label{eq:alphas_compact}
\end{equation}
where $\tilde{S}_{\ell}(k) = S_{\ell}(k)G_{\ell}^{(d)}(k)$ is the source image at the origin and $\talpha = 4 \pi i^{n}Y_{n m}^{*}\left(\hat{\boldsymbol{x}}_{\ell}\right)$ are source-independent coefficients that are only a function of the source direction. The coefficients $\alpha_{n m}$ can be estimated using the signals acquired by a spherical microphone array with $Q$ capsules as 
\begin{equation}\label{eq:mic_encoded}
\alpha_{n m}(k) 
\approx \frac{1}{b_{n}(k r)} \sum_{q=1}^{Q} w_{q} P\left(\boldsymbol{x}_{q}^{\prime}, k\right) Y_{n m}^{*}\left(\xu\right),
\end{equation}
where $w_{q} \forall q$ are weights that ensure the validity of the orthonormal property of the SHs. The source localization problem in the SHD consists in the estimation of the DOA vectors $\xlu$ corresponding to the active sound sources from the measured SH coefficients in Eq.~\eqref{eq:mic_encoded}. The geometry of the problem is graphically depicted in Fig.\ref{fig:setup}.

\section{METHOD}\label{sec:method}%
Consider the matrix arrangement of the SH coefficients extracted for a set of frequencies $\mathcal{F}=\left\{k_1, k_2, \dots,k_F \right\}$
at a given time frame $\tau$:
\begin{equation}
    \mathbf{A}_\tau =  \left[\mathbf{a}_N(k_1), \mathbf{a}_N(k_2),\dots,\mathbf{a}_N(k_F)\right]\in \mathbb{C}^{C \times F},
    \label{eq:Amatrix0}
\end{equation}
where each column $\mathbf{a}_N(k)$ contains the coefficients extracted for a given frequency $k$ up to the $N$th order, i.e.
\begin{equation}
    \mathbf{a}_N(k) = \left[\alpha_{0,0}(k),\alpha_{-1,1}(k),\dots,\alpha_{NN}(k) \right]^T.
\end{equation}
By writing the contribution from the $\ell$th source as:
\begin{equation}
    \mathbf{s}_{\ell} = \left[\tilde{S}_{\ell}(k_1),\, \tilde{S}_{\ell}(k_2),\dots,\tilde{S}_{\ell}(k_F) \right]^T,
\end{equation}
and using Eq.~\eqref{eq:alphas_compact}, matrix $\mathbf{A}_\tau$ can be written as a sum of rank-1 matrices:
\begin{equation}
    \mathbf{A}_\tau = \sum_{\ell=1}^{L} \balpha \mathbf{s}_{\ell}^{T},
    \label{eq:Amatrix}
\end{equation}
where the vector $\balpha \in \mathbb{C}^{C}$ compiles all the source-independent coefficients up to the $N$th order, i.e.
\begin{equation}
    \balpha\coloneqq\left[\tilde{\alpha}_{00},\, \tilde{\alpha}_{1 -1}, \cdots, \tilde{\alpha}_{nm}, \cdots, \tilde{\alpha}_{NN}\right]^{T}.
\end{equation}
We denote this vector the modal directional pattern (MDP) of the $\ell$th source (the dependence of its elements on the DOA $\xlu$ has been omitted for notation simplicity). Therefore, under an ideal noiseless anechoic case, it holds that $\mathrm{rank}(\mathbf{A}_\tau)~=~L_a$, where $L_a\leq L$ is the number of active sources at the analyzed frame. The objective of the method is to find a low-rank approximation of $\mathbf{A}_\tau$ that leads to an estimate of the MDP of the primary active source, mapping such MDP to the most likely DOA vector.
\subsection{Proposed Approach}%
\subsubsection{Step 1: Normalization}%
The norm of the MDP is DOA-independent. In particular
\begin{equation}
    \|\balpha\| = \balphan = \sqrt{4\pi}(N+1) \quad \forall \xlu.
    \label{eq:mdp_norm}
\end{equation}
Under the assumption that each column of $\mathbf{A}_\tau$ is dominated by a single source (W-disjoint orthogonality), we exploit the property in Eq.~\eqref{eq:mdp_norm} to normalize $\mathbf{A}_\tau$ column-wise.
Indeed, by assuming $L~=~1$, it holds that
$\alpha_{n m}(k)~=~ \talphad\tilde{S}_{\check{\ell}}(k)$ and, thus, $ \mathbf{a}_N(k)~=~  {S}_{\check{\ell}}(k)\balphad$, where the sub-index $\check{\ell}$ indicates the primary active source. According to Eq.~\eqref{eq:mdp_norm}
\begin{equation}
\|\mathbf{a}_N(k)\| = |{S}_{\check{\ell}}(k)| \sqrt{4\pi}(N+1),
\label{eq:mode_norm}
\end{equation}
thus, we normalize the coefficients as follows:
\begin{equation}
    \overline{\alpha}_{nm}(k) = |\hat{S}_{\check{\ell}}(k)| \frac{\alpha_{nm}(k)}{\|\mathbf{c}_n\|}\sqrt{4\pi (2n+1)},
\end{equation}
\begin{equation}
    \mathbf{c}_n\coloneqq\left[\alpha_{n,-n}(k), \cdots, \alpha_{nn}(k)\right] \in \mathbb{C}^{2n+1},
\end{equation}
where
\begin{equation}
    |\hat{S}_{\check{\ell}}(k)| = \frac{1}{N}\sum_{n=0}^{N} \frac{\|\mathbf{c}_n\|}{\sqrt{4\pi(2n+1)}}
\end{equation}
is a mean-based estimate of the magnitude of the primary source. Similarly to Eq.~\eqref{eq:Amatrix0}, the resulting normalized vectors, $\overline{\mathbf{a}}_N$, lead to the normalized matrix $\An$.
\begin{figure*}[t]%
  \begin{subfigure}[]{%
  \input{images/pd_bar_0_rev}\label{fig:pd_0}%
  }%
  \end{subfigure}%
  \begin{subfigure}[]{%
  \input{images/pd_bar_8_rev}\label{fig:pd_05}%
  }%
  \end{subfigure}%
  \begin{subfigure}[]{%
  \input{images/pd_bar_9_rev}\label{fig:pd_1}%
  }%
  \end{subfigure}
  \begin{subfigure}[]{%
  \input{images/rmse_bar_0_rev}\label{fig:rmse_0}%
  }%
  \end{subfigure}%
  \begin{subfigure}[]{%
  \input{images/rmse_bar_8_rev}\label{fig:rmse_05}%
  }%
  \end{subfigure}%
  \begin{subfigure}[]{%
  \input{images/rmse_bar_9_rev}\label{fig:rmse_1}%
  }%
  \end{subfigure}%
\caption{Probability of detection ($\PD$) and DOA $\RMSE$ for different simulated conditions. Left column (a, d): $T_{60}=\SI{0.0}{\second}$. Middle column (b, e): $T_{60}=\SI{0.5}{\second}$. Right column (c, f): $T_{60}=\SI{1}{\second}$.}%
\label{fig:res}%
\end{figure*}
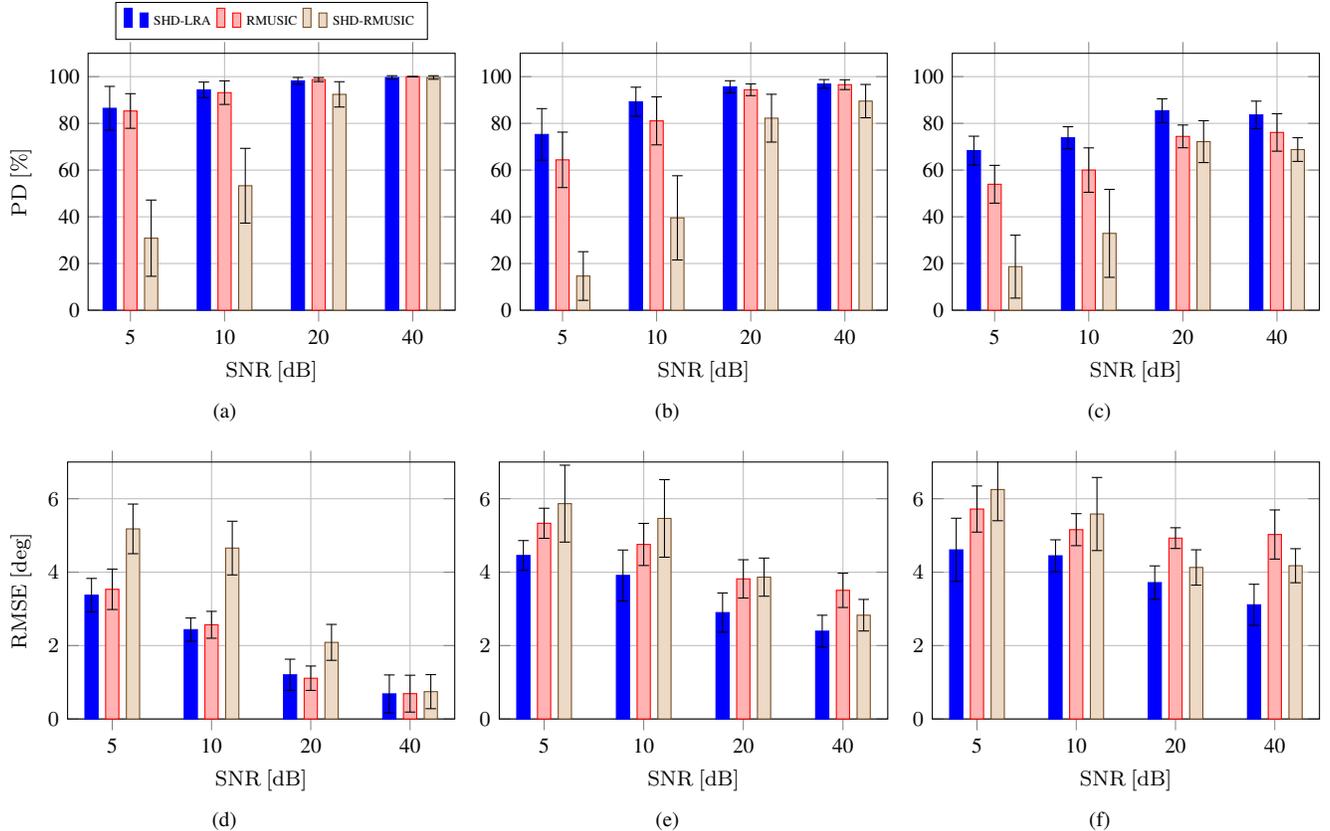%
\subsubsection{Step 2: Singular Value Decomposition}%
A low-rank approximation of $\An$ can be obtained by solving
\begin{equation}
    \min_{\hat{\mathbf{A}}_\tau} \quad \| \An-\hat{\mathbf{A}}_\tau\|_F, \quad \textrm{subject to} \quad \textrm{rank}(\hat{\mathbf{A}}_\tau)\leq r
\end{equation}
where $r$ is the rank of the approximating matrix $\hat{\mathbf{A}}_\tau$, and $\|~\cdot~\|_{F}$ denotes the Frobenius norm. The problem has analytic solution  in terms of the singular value decomposition (SVD) of $\An$. Let us factorize $\An$ as
\begin{equation}
    \An = \mathbf{U}\mathbf{\Sigma}\mathbf{V}^H,
\end{equation}
where $\mathbf{U}~\in~\mathbb{C}^{C \times C}$ is the matrix containing the left singular vectors, $\mathbf{\Sigma}~\in ~\mathbb{R}^{C \times F}$ is the diagonal matrix containing the ordered singular values and $\mathbf{V}~\in~ \mathbb{C}^{F \times F}$ is the matrix containing the right singular vectors. The particular rank-$1$ matrix that best approximates $\An$ is given by
\begin{equation}
    \hat{\mathbf{A}}_\tau=\sigma_1\mathbf{u}_1 \mathbf{v}_1^{T} \approx \balphan\| \mathbf{s}_{\check{\ell}}\|\frac{\balphad}{\balphan}\frac{\mathbf{s}_{\check{\ell}}^T}{\| \mathbf{s}_{\check{\ell}}\|},
    \label{eq:Asvd}
\end{equation}
where $\mathbf{u}_1$ and $\mathbf{v}_1$ are the first columns of the corresponding SVD matrices and $\sigma_1$ is the first ordered singular value. Therefore, an estimate of the primary MDP is directly given by using Eqs.~\eqref{eq:Asvd} and \eqref{eq:mdp_norm} as:
\begin{equation}
    \hbalpha =  \balphan \mathbf{u}_1 = \sqrt{4\pi}(N+1)\mathbf{u}_1.
\end{equation}

The $\hbalpha$ retrieved from the first left singular vector is assumed to be significantly robust to noise, as it comes from the best rank-1 approximation of $\An$ in the least-squares provided by the SVD.
\subsubsection{Step 3: DOA mapping}%
The final DOA estimate is obtained through pattern matching by using a pre-computed dictionary of MDPs corresponding to a predefined spatial grid on the unit sphere surface:
\begin{equation}
    \doublehat{\mathbf{x}}_\ell = \min_{\hat{\mathbf{x}}_\ell \in \mathcal{D}} \|\hbalpha - \balpha \|^2,
\end{equation}
where $\mathcal{D}$ is the set containing the considered candidate locations.
\section{EXPERIMENTS}%
\label{sec:experiments}%
The performance of the proposed approach was evaluated over an extensive set of synthetically generated recordings with random source-receiver configurations using the image-source method \cite{allen1979image} in a shoe-box room with dimensions $\SI{10}{\meter}$ $\times$ $\SI{8}{\meter}$ $\times$ $\SI{6}{\meter}$. 
To assess the effect of noise and reverberation, we ran simulations both for anechoic ($T_{60}=\SI{0}{\second}$) and reverberant conditions ($T_{60} = \left\{0.5,\, 1\right\}\si{\second}$) considering a range of Signal-to-Noise Ratio (SNR) values ($\textrm{SNR} = \left\{5,\, 10,\, 20,\, 40 \right\}\si{\decibel}$), with 10 runs per room condition. The simulations included the theoretical response of a rigid spherical microphone array with 32 channels, hence the maximum SH order is $N=4$. 
The sources were randomly placed in the room with a distance from the array of $\SI{2}{\meter}$. We processed the signals at a sampling rate of $\SI{8}{kHz}$, the STFT adopts a hamming window of size 512 with $\SI{50}{\percent}$ overlap and 512 FFT points. The sources were randomly taken in each simulation run  from a set of 3 male and 3 female Japanese and English speakers extracted from \cite{ono20132013}.

We compare the performance of the proposed SHD-LRA method with respect to the recent MUSIC-based models RMUSIC and SHD-RMUSIC, both introduced in \cite{hu2020multiple}.
Similarly to \cite{hu2020multiple}, we perform the analysis over a limited frequency range ($1$-$\SI{2.5}{\kilo\hertz}$), hence exploiting the SHs up to the $3$rd order. All the methods performed localization by considering the same spatial grid of candidate DOAs, with a separation of 3 degrees in elevation and 2 degrees in azimuth. Location estimates are obtained over signal frames having a duration of 0.3 s. To avoid undesired
effects in the performance analysis due to speech silences, the source
signals were manually processed to keep only audio segments with
speech activity. 
\subsection{Metrics}
The performance metrics used in the evaluation are the probability of detection ($\PD$) and the DOA root mean squared error ($\RMSE$), the former computed as the percentage of DOA estimates below an absolute DOA error of 10 degrees. 
Given the angular error for a given DOA estimate $\doublehat{\mathbf{x}}_\ell$:
\begin{equation}
    \psi_e = \arccos \left( \doublehat{\mathbf{x}}_\ell ^{T} \hat{\mathbf{x}}_\ell\right),
\end{equation}
we define nonanomalous estimates as those where $|\psi_e| < \SI{10}{\degree}$. The $\PD$ is obtained as
\begin{equation}
    \PD = \frac{N_a}{N_T},
\end{equation}
where $N_a$ is the number of nonanomalous estimates and $N_T$ the total number of estimates. The DOA RMSE is computed as:
\begin{equation}
    \RMSE = \sqrt{\frac{1}{N_a}\sum_{i\in \mathcal{N}_a}\psi^2_{e,i}},
\end{equation}
where $\mathcal{N}_a$ is the set containing the nonanomalous DOA estimates.
\subsection{Discussion}%
The results are collected in Fig.~\ref{fig:res}, which shows the average and standard deviation of the considered metrics computed across the  different simulation runs. The bar graphs in (a, b, c) show clearly how, in general, the $\PD$ is significantly affected by the additive noise level and room reverberation. As expected, while all methods provide almost perfect detection performance in anechoic and high SNR conditions, the performance is severely degraded at lower SNRs or under stronger room reflections. The proposed method SHD-LRA is, however, the one showing a more robust behavior. While both RMUSIC and SHD-LRA provide almost identical results for (a) $T_{60}=0$ s, the performance drop is higher for RMUSIC when increasing the reverberation time (b, c), especially at lower SNRs. In all cases, the SHD-RMUSIC method showed poorer performance, with a remarkable sensitivity to noise compared to the other two methods. 

A similar trend is observed for the DOA RMSE in the graphs of the bottom panel (d, e, f). While, in anechoic conditions (d), the performance of SHD-LRA is quite similar to that of RMUSIC, a general improvement is observed for the proposed method in reverberant conditions (e, f) for every SNR. Interestingly, when reverberation is present and the SNR is high, the angular error of the SHD-RMUSIC method is smaller than that of RMUSIC.
However, the low $\PD$ shown by SHD-RMUSIC indicates that the overall performance is less robust to non-ideal conditions.
In general, with moderate and high reverberation the proposed method SHD-LRA outperforms the other two baseline approaches.

Finally, note that typical MUSIC-based methods need to compute an SVD for the covariance matrix of each frequency bin, which is of size $Q \times Q$. In our method, we perform one single SVD for all the considered frequencies at once from a matrix of size $C \times F$. Then, the cost of MUSIC-based methods is in the order of $\mathcal{O}(FQ^3)$, while that of SHD-LRA is $\mathcal{O}(FC^2)$, which is significantly lower as $C\leq Q$.

\section{CONCLUSION}\label{sec:conclusion}%
In this paper, we have presented a method for acoustic source localization working in the spherical harmonics domain. The method is based on the extraction of the underlying modal directional pattern (MDP) corresponding to the direction of arrival (DOA) of the source. To this end, the spherical harmonics coefficients computed from the acquired multichannel signal at multiple time-frequency points, are normalized and arranged into a matrix that is assumed to be low-rank under ideal conditions. While the normalization step acts as a denoising stage, the SVD applied over the resulting matrix helps to identify the target dominant MDP and, consequently, the DOA of the source. An evaluation comparing to other recent approaches using simulated data confirms the robustness and potential of the proposed approach. Future work will consider the extension of the method to multi-source localization and a more comprehensive evaluation involving recordings acquired in real scenarios.
\bibliographystyle{ieeetr}
\bibliography{strings,refs}
\end{document}

%% file: images/pd_bar_0_rev.tex
\begin{tikzpicture}%
	\begin{axis}[%
	ybar=.1cm,%
    bar width=5pt,%
    width = 0.363\textwidth,
    height = \figSize,%
    grid = both,%
    enlarge x limits=0.15,%
    ymin = 0,%
    ymax = 1.1,%
    major grid style = {lightgray},%
    minor grid style = {lightgray!25},%
    ylabel = $\PD$,%
    y unit= \percent,%
    xmode=log,%
    log ticks with fixed point,%
    xtick = {5,10,20,40},%
    xticklabels = {5,10,20,40},%
    ytick = {0, 0.2, 0.4, 0.6, 0.8, 1},%
    yticklabels = {0, 20,40,60,80,100},%
    x unit = \decibel,%
    xlabel={$\operatorname{SNR}$},%
    style={ font=\footnotesize},%
    legend style={at={(0.5,1.2)},%
    anchor=north,legend columns=3, font=\tiny},%
    ]%
    \addplot+[%
        blue, %
        error bars/.cd, %
        y dir=both, %
        y explicit, %
        error bar style=black%
        ] %
		table[x index=0, y index=1, y error index=2] {images/data/pd_shd_lra_0.txt};%
	\addlegendentry{SHD-LRA};%
	\addplot+[%
        error bars/.cd, %
        y dir=both, %
        y explicit,%
        error bar style=black%
        ] %
		table[x index=0, y index=1, y error index=2] {images/data/pd_rmusic_0.txt};%
	\addlegendentry{RMUSIC};%
	\addplot+[%
        error bars/.cd, %
        y dir=both, %
        y explicit,%
        error bar style=black%
        ] %
		table[x index=0, y index=1, y error index=2] {images/data/pd_shd_rmusic_0.txt};%
	\addlegendentry{SHD-RMUSIC};%
\end{axis}%
\end{tikzpicture}%

%% file: images/pd_bar_8_rev.tex
\begin{tikzpicture}
	\begin{axis}[%
	ybar=.1cm,
    bar width=5pt,%
    width = 0.363\textwidth,
    height = \figSize,%
    grid = both,%
    enlarge x limits=0.25,%
    enlarge x limits=0.15,%
    ymin = 0,%
    ymax = 1.1,%
    major grid style = {lightgray},%
    minor grid style = {lightgray!25},%
    xmode=log,%
    log ticks with fixed point,%
    xtick = {5,10,20,40},%
    xticklabels = {5,10,20,40},%
    ytick = {0,0.2,0.4,0.6,0.8,1},%
    yticklabels = {0, 20,40,60,80,100},%
    x unit = \decibel,%
    xlabel={$\operatorname{SNR}$},%
    style={ font=\footnotesize},%
    legend style={at={(0.5,1.5)},%
    anchor=north,legend columns=3, font=\tiny},%
    ]%
    \addplot+[%
        blue, %
        error bars/.cd, %
        y dir=both, %
        y explicit, %
        error bar style=black%
        ] %
		table[x index=0, y index=1, y error index=2] {images/data/pd_shd_lra_8.txt};%
	\addplot+[%
        error bars/.cd, %
        y dir=both, %
        y explicit,%
        error bar style=black%
        ] %
		table[x index=0, y index=1, y error index=2] {images/data/pd_rmusic_8.txt};%
	\addplot+[%
        error bars/.cd, %
        y dir=both, %
        y explicit,%
        error bar style=black%
        ] %
		table[x index=0, y index=1, y error index=2] {images/data/pd_shd_rmusic_8.txt};%
\end{axis}%
\end{tikzpicture}%

%% file: images/pd_bar_9_rev.tex
\begin{tikzpicture}%
	\begin{axis}[%
	ybar=.1cm,%
    bar width=5pt,%
    width = 0.363\textwidth,
    height = \figSize,%
    grid = both,%
    enlarge x limits=0.25,%
    enlarge x limits=0.15,%
    ymin = 0,%
    ymax = 1.1,%
    major grid style = {lightgray},%
    minor grid style = {lightgray!25},%
    xmode=log,%
    log ticks with fixed point,%
    xtick = {5,10,20,40},%
    xticklabels = {5,10,20,40},%
    ytick = {0,0.2,0.4,0.6,0.8,1},%
    yticklabels = {0, 20,40,60,80,100},%
    x unit = \decibel,%
    xlabel={$\operatorname{SNR}$},%
    style={ font=\footnotesize},%
    legend style={at={(0.5,1.5)},%
    anchor=north,legend columns=3, font=\tiny},%
    ]%
    \addplot+[%
        blue, %
        error bars/.cd, %
        y dir=both, %
        y explicit, %
        error bar style=black%
        ] %
		table[x index=0, y index=1, y error index=2] {images/data/pd_shd_lra_9.txt};%
	\addplot+[%
        error bars/.cd,%
        y dir=both, %
        y explicit,%
        error bar style=black%
        ] %
		table[x index=0, y index=1, y error index=2] {images/data/pd_rmusic_9.txt};%
	\addplot+[%
        error bars/.cd, %
        y dir=both, %
        y explicit,%
        error bar style=black%
        ] %
		table[x index=0, y index=1, y error index=2] {images/data/pd_shd_rmusic_9.txt};%
\end{axis}%
\end{tikzpicture}%

%% file: images/rmse_bar_0_rev.tex
\begin{tikzpicture}
	\begin{axis}[
   	ybar=.1cm,
    bar width=5pt,
    width = 0.378\textwidth,
    height = \figSize,
    grid = both,%
    enlarge x limits=0.15,%
    ymin = 0,%
    ymax = 7,%
    major grid style = {lightgray},%
    minor grid style = {lightgray!25},%
    ylabel = $\RMSE$,%
    y unit= \deg,%
    xmode=log,%
    log ticks with fixed point,%
    xtick = {5,10,20,40},%
    xticklabels = {5,10,20,40},%
    x unit = \decibel,%
    xlabel={$\operatorname{SNR}$},%
    style={ font=\footnotesize},%
    legend style={at={(0.5,1.3)},%
    anchor=north,legend columns=3, font=\tiny},%
    ]
    \addplot+[
        blue, 
        error bars/.cd, 
        y dir=both, 
        y explicit,
        error bar style=black
        ] %
		table[x index=0, y index=1, y error index=2] {images/data/doa_shd_lra_0.txt};%
	\addplot+[
        error bars/.cd, 
        y dir=both, 
        y explicit,
        error bar style=black
        ] %
		table[x index=0, y index=1, y error index=2] {images/data/doa_rmusic_0.txt};%
	\addplot+[
        error bars/.cd, 
        y dir=both, 
        y explicit,
        error bar style=black
        ] %
		table[x index=0, y index=1, y error index=2] {images/data/doa_shd_rmusic_0.txt};%
\end{axis}%
\end{tikzpicture}%

%% file: images/rmse_bar_8_rev.tex
\begin{tikzpicture}%
	\begin{axis}[%
   	ybar=.1cm,%
    bar width=5pt,%
    width = 0.379\textwidth,
    height = \figSize,%
    grid = both,%
    ymin=0,%
    ymax=7,%
    enlarge x limits=0.15,%
    major grid style = {lightgray},%
    minor grid style = {lightgray!25},%
    xmode=log,%
    log ticks with fixed point,%
    xtick = {5,10,20,40},%
    xticklabels = {5,10,20,40},%
    x unit = \decibel,%
    xlabel={$\operatorname{SNR}$},%
    style={ font=\footnotesize},%
    legend style={at={(0.5,1.3)},%
    anchor=north,legend columns=3, font=\tiny},%
    ]%
    \addplot+[%
        blue, %
        error bars/.cd, %
        y dir=both, %
        y explicit,%
        error bar style=black%
        ] %
		table[x index=0, y index=1, y error index=2] {images/data/doa_shd_lra_8.txt};%
	\addplot+[%
        error bars/.cd, %
        y dir=both, %
        y explicit,%
        error bar style=black%
        ] %
		table[x index=0, y index=1, y error index=2] {images/data/doa_rmusic_8.txt};%
	\addplot+[%
        error bars/.cd, %
        y dir=both, %
        y explicit,%
        error bar style=black%
        ] %
		table[x index=0, y index=1, y error index=2] {images/data/doa_shd_rmusic_8.txt};%
\end{axis}%
\end{tikzpicture}%

%% file: images/rmse_bar_9_rev.tex
\begin{tikzpicture}%
	\begin{axis}[%
   	ybar=.1cm,%
    bar width=5pt,%
    width = 0.378\textwidth,
    height = \figSize,
    grid = both,%
    ymin=0,%
    ymax=7,%
    enlarge x limits=0.15,%
    major grid style = {lightgray},%
    minor grid style = {lightgray!25},%
    xmode=log,%
    log ticks with fixed point,%
    xtick = {5,10,20,40},%
    xticklabels = {5,10,20,40},%
    x unit = \decibel,%
    xlabel={$\operatorname{SNR}$},%
    style={ font=\footnotesize},%
    legend style={at={(0.5,1.3)},%
    anchor=north,legend columns=3, font=\tiny},%
    ]%
    \addplot+[%
        blue, %
        error bars/.cd, %
        y dir=both, %
        y explicit,%
        error bar style=black%
        ] %
		table[x index=0, y index=1, y error index=2] {images/data/doa_shd_lra_9.txt};%
	\addplot+[%
        error bars/.cd, %
        y dir=both, %
        y explicit,%
        error bar style=black%
        ] %
		table[x index=0, y index=1, y error index=2] {images/data/doa_rmusic_9.txt};%
	\addplot+[%
        error bars/.cd, %
        y dir=both, %
        y explicit,%
        error bar style=black%
        ] %
		table[x index=0, y index=1, y error index=2] {images/data/doa_shd_rmusic_9.txt};%
\end{axis}%
\end{tikzpicture}%

%% file: Template.bbl
\begin{thebibliography}{10}

\bibitem{cobos2017survey}
M.~Cobos, F.~Antonacci, A.~Alexandridis, A.~Mouchtaris, and B.~Lee, ``A survey
  of sound source localization methods in wireless acoustic sensor networks,''
  {\em Wireless Comm. and Mobile Computing}, vol.~2017.

\bibitem{gannot2017consolidated}
S.~Gannot, E.~Vincent, S.~Markovich-Golan, and A.~Ozerov, ``A consolidated
  perspective on multimicrophone speech enhancement and source separation,''
  {\em IEEE Trans. on audio, speech, and language Process.}, vol.~25, no.~4,
  pp.~692--730, 2017.

\bibitem{markovic2016extraction}
D.~{Marković}, F.~{Antonacci}, L.~{Bianchi}, S.~{Tubaro}, and A.~{Sarti},
  ``Extraction of acoustic sources through the processing of sound field maps
  in the ray space,'' {\em IEEE Trans. on audio, speech, and language
  Process.}, vol.~24, pp.~2481--2494, Dec 2016.

\bibitem{pezzoli2020parametric}
M.~Pezzoli, F.~Borra, F.~Antonacci, S.~Tubaro, and A.~Sarti, ``A parametric
  approach to virtual miking for sources of arbitrary directivity,'' {\em IEEE
  Trans. on audio, speech, and language Process.}, vol.~28, pp.~2333--2348,
  2020.

\bibitem{cobos2022overview}
M.~Cobos, J.~Ahrens, K.~Kowalczyk, and A.~Politis, ``An overview of machine
  learning and other data-based methods for spatial audio capture, processing,
  and reproduction,'' {\em EURASIP Journal on Audio, Speech, and Music
  Process.}, vol.~2022, no.~1, pp.~1--21, 2022.

\bibitem{yao2002maximum}
K.~Yao, J.~C. Chen, and R.~E. Hudson, ``Maximum-likelihood acoustic source
  localization: experimental results,'' in {\em Int. Conf. on Acoust., Speech
  and signal Process. (ICASSP)}, vol.~3, pp.~III--2949, IEEE, 2002.

\bibitem{dibiase2001robust}
J.~H. DiBiase, H.~F. Silverman, and M.~S. Brandstein, ``Robust localization in
  reverberant rooms,'' in {\em Microphone arrays}, pp.~157--180, Springer,
  2001.

\bibitem{cobos2010modified}
M.~Cobos, A.~Marti, and J.~J. Lopez, ``A modified srp-phat functional for
  robust real-time sound source localization with scalable spatial sampling,''
  {\em IEEE Signal Process. Letters}, vol.~18, no.~1, pp.~71--74, 2010.

\bibitem{benesty2000adaptive}
J.~Benesty, ``Adaptive eigenvalue decomposition algorithm for passive acoustic
  source localization,'' {\em J. Acoust. Soc. Am.}, vol.~107, no.~1,
  pp.~384--391, 2000.

\bibitem{stoica1989music}
P.~Stoica and A.~Nehorai, ``Music, maximum likelihood, and cramer-rao bound,''
  {\em IEEE Trans. on audio, speech, and language Process.}, vol.~37, no.~5,
  pp.~720--741, 1989.

\bibitem{jo2018direction}
B.~Jo and J.-W. Choi, ``Direction of arrival estimation using nonsingular
  spherical esprit,'' {\em J. Acoust. Soc. Am.}, vol.~143, no.~3,
  pp.~EL181--EL187, 2018.

\bibitem{markovic2013soundfield}
D.~Markovic, F.~Antonacci, A.~Sarti, and S.~Tubaro, ``Soundfield imaging in the
  ray space,'' {\em IEEE Trans. on audio, speech, and language Process.},
  vol.~21, no.~12, pp.~2493--2505, 2013.

\bibitem{williams1999fourier}
E.~G. Williams, {\em Fourier acoustics: sound radiation and nearfield
  acoustical holography}.
\newblock Elsevier, 1999.

\bibitem{pezzoli2021ray}
M.~Pezzoli, J.~J. Carabias-Orti, M.~Cobos, F.~Antonacci, and A.~Sarti,
  ``Ray-space-based multichannel nonnegative matrix factorization for audio
  source separation,'' {\em IEEE Signal Process. Letter}, vol.~28,
  pp.~369--373, 2021.

\bibitem{mitsufuji2020multichannel}
Y.~Mitsufuji, N.~Takamune, S.~Koyama, and H.~Saruwatari, ``Multichannel blind
  source separation based on evanescent-region-aware non-negative tensor
  factorization in spherical harmonic domain,'' {\em Trans. on audio, speech,
  and language Process.}, vol.~29, pp.~607--617, 2020.

\bibitem{pezzoli2022sparsity}
M.~Pezzoli, M.~Cobos, F.~Antonacci, and A.~Sarti, ``Sparsity-based sound field
  separation in the spherical harmonics domain,'' in {\em Int. Conf. on
  Acoust., Speech and signal Process. (ICASSP)}, pp.~1051--1055, IEEE, 2022.

\bibitem{fahim2017sound}
A.~Fahim, P.~N. Samarasinghe, and T.~D. Abhayapala, ``Sound field separation in
  a mixed acoustic environment using a sparse array of higher order spherical
  microphones,'' in {\em Hands-free Speech Commun. and Microphone Arrays},
  pp.~151--155, IEEE, 2017.

\bibitem{bianchi_ray_2016}
L.~Bianchi, F.~Antonacci, A.~Sarti, and S.~Tubaro, ``The ray space transform: A
  new framework for wave field processing,'' {\em IEEE Transactions on Signal
  Processing}, vol.~64, pp.~5696--5706, Nov. 2016.

\bibitem{abhayapala2003coherent}
T.~D. Abhayapala and H.~Bhatta, ``Coherent broadband source localization by
  modal space processing,'' in {\em 10th Int. Conf. on Telecomm.}, vol.~2,
  pp.~1617--1623, IEEE, 2003.

\bibitem{teutsch2005eigen}
H.~Teutsch and W.~Kellermann, ``Eigen-beam processing for direction-of-arrival
  estimation using spherical apertures,'' in {\em Hands-free Speech Commun. and
  Microphone Arrays}, vol.~4, IEEE, 2005.

\bibitem{hu2019sound}
Y.~Hu, P.~N. Samarasinghe, and T.~D. Abhayapala, ``Sound source localization
  using relative harmonic coefficients in modal domain,'' in {\em Workshop
  Appl. Signal Process. Audio Acoust.}, pp.~348--352, IEEE, 2019.

\bibitem{hu2020unsupervised}
Y.~Hu, P.~N. Samarasinghe, T.~D. Abhayapala, and S.~Gannot, ``Unsupervised
  multiple source localization using relative harmonic coefficients,'' in {\em
  Int. Conf. on Acoust., Speech and signal Process. (ICASSP)}, pp.~571--575,
  IEEE, 2020.

\bibitem{hu2021evaluation}
Y.~Hu, P.~N. Samarasinghe, S.~Gannot, and T.~D. Abhayapala, ``Evaluation and
  comparison of three source direction-of-arrival estimators using relative
  harmonic coefficients,'' in {\em Int. Conf. on Acoust., Speech and signal
  Process. (ICASSP)}, pp.~815--819, IEEE, 2021.

\bibitem{hu2021decoupled}
Y.~Hu, T.~D. Abhayapala, P.~N. Samarasinghe, and S.~Gannot, ``Decoupled
  direction-of-arrival estimations using relative harmonic coefficients,'' in
  {\em 28th Eur. Signal Process. Conf.}, pp.~246--250, IEEE, 2021.

\bibitem{hu2022decoupled}
Y.~Hu, P.~N. Samarasinghe, S.~Gannot, and T.~D. Abhayapala, ``Decoupled
  multiple speaker direction-of-arrival estimator under reverberant
  environments,'' {\em Trans. on audio, speech, and language Process.}, 2022.

\bibitem{hu2022closed}
Y.~Hu and S.~Gannot, ``Closed-form single source direction-of-arrival estimator
  using first-order relative harmonic coefficients,'' in {\em Int. Conf. on
  Acoust., Speech and signal Process. (ICASSP)}, pp.~726--730, IEEE, 2022.

\bibitem{hu2020multiple}
Y.~Hu, T.~D. Abhayapala, and P.~N. Samarasinghe, ``Multiple source direction of
  arrival estimations using relative sound pressure based music,'' {\em Trans.
  on audio, speech, and language Process.}, vol.~29, pp.~253--264, 2020.

\bibitem{ward2001reproduction}
D.~B. Ward and T.~D. Abhayapala, ``Reproduction of a plane-wave sound field
  using an array of loudspeakers,'' {\em Trans. on audio, speech, and language
  Process.}, vol.~9, no.~6, pp.~697--707, 2001.

\bibitem{allen1979image}
J.~B. Allen and D.~A. Berkley, ``Image method for efficiently simulating
  small-room acoustics,'' {\em The Journal of the Acoustical Society of
  America}, vol.~65, no.~4, pp.~943--950, 1979.

\bibitem{ono20132013}
N.~Ono, Z.~Koldovsk{\`y}, S.~Miyabe, and N.~Ito, ``The 2013 signal separation
  evaluation campaign,'' in {\em International workshop on machine learning for
  signal processing (MLSP)}, pp.~1--6, IEEE, 2013.

\end{thebibliography}
